\documentclass[aps,prd,twocolumn,nofootinbib,showpacs,floatfix,preprintnumbers,groupedaddress]{revtex4}
\usepackage{graphicx}
\usepackage{epsfig}

\newcommand{\Sla}[1]{/\!\!\!\!#1}
\def\lsim{\raise0.3ex\hbox{$\;<$\kern-0.75em\raise-1.1ex\hbox{$\sim\;$}}}
\def\gsim{\raise0.3ex\hbox{$\;>$\kern-0.75em\raise-1.1ex\hbox{$\sim\;$}}}

\def\hbar{\hspace{0pt}\raisebox{1pt}{$-$} \hspace{-7pt} h}

\newcommand {\ignore}[1]{}

\begin{document}
\preprint{YITP-SB-11-04}

\title{Determination of the Spin of New Resonances \\
in Electroweak Gauge Boson Pair Production at the LHC}

\author{O.\ J.\ P.\ \'Eboli}
\email{eboli@fma.if.usp.br}
\affiliation{Instituto de F\'{\i}sica,
             Universidade de S\~ao Paulo, S\~ao Paulo -- SP, Brazil.}

\author{Chee Sheng Fong}
\affiliation{%
  C.N.~Yang Institute for Theoretical Physics, SUNY at Stony Brook,
  Stony Brook, NY 11794-3840, USA}

\author{J.\ Gonzalez-Fraile}
\email{fraile@ecm.ub.es}
\affiliation{%
  Departament d'Estructura i Constituents de la Mat\`eria and
  ICC-UB, Universitat de Barcelona, 647 Diagonal, E-08028 Barcelona,
  Spain}

\author{M.\ C.\ Gonzalez--Garcia} \email{concha@insti.physics.sunysb.edu}
\affiliation{%
  C.N.~Yang Institute for Theoretical Physics, SUNY at Stony Brook,
  Stony Brook, NY 11794-3840, USA}
\affiliation{%
  Instituci\'o Catalana de Recerca i Estudis Avan\c{c}ats (ICREA),
  Departament d'Estructura i Constituents de la Mat\`eria, Universitat
  de Barcelona, 647 Diagonal, E-08028 Barcelona, Spain}


\begin{abstract}
\vspace*{1cm}

The appearance of spin-1 resonances associated to the electroweak
symmetry breaking (EWSB) sector is expected in many extensions of the
Standard Model. We analyze the CERN Large Hadron Collider potential to
probe the spin of possible new charged and neutral vector resonances
through the purely leptonic processes $pp \to Z^\prime \to \ell^+
\ell^{\prime -} \, \Sla{E}_T, $ and $ pp \to W^\prime \to \ell^{\prime
  \pm} \ell^+ \ell^- \,\Sla{E}_T, $ with $\ell, \ell^\prime = e$ or
$\mu$. We perform a model independent analysis and demonstrate that
the spin of the new states can be determined with 99\% CL in a large
fraction of the parameter space where these resonances can be observed
with 100 fb$^{-1}$. We show that the best sensitivity to the spin is
obtained by directly studying correlations between the final state
leptons, without the need of reconstructing the events in their
center--of--mass frames.

\end{abstract}

\maketitle



\section{Introduction}

One of the prime objectives of the CERN Large Hadron Collider (LHC) is
to probe directly the EWSB sector.  The analysis of partial wave
unitarity of longitudinal weak boson scattering guarantees that there
should exist a new state at the TeV scale or that this process becomes
strongly interacting at high energies~\cite{Lee:1977yc,Lee:1977eg}. In
several extensions of the Standard Model (SM) the new resonances
associated to the unitarity restoration are expected to have spin
1. For instance, in Higgsless models a tower of spin--1 particles is
responsible for cutting off the growth of the electroweak gauge boson
scattering amplitude without the presence of any scalar (Higgs)
field~\cite{Csaki:2003dt}. Another attractive possibility is that the
electroweak symmetry breaking is associated to a new strongly
interacting sector~\cite{TC}. These models also exhibit new vector
states that contribute to the unitarization of the weak gauge boson
scattering~\cite{NTC}.  \smallskip

Thus, a common feature of many EWSB scenarios, as the ones above
mentioned, is the existence of new vector resonances, $Z'$ and $W'$,
that couple to $W^+W^-$ and $W^\pm Z$ pairs respectively. But,
generically, their properties, such as mass, width, and couplings to
SM particles, are model dependent.  In this respect, the {\sl model
  independent} channels for detection of such spin-1 resonances would
be their production via weak boson fusion (WBF) or its associated
production with an electroweak gauge boson, since both processes only
involve their couplings to electroweak gauge bosons.  Unfortunately
for a $Z'$ these signals are unobservable in a clean purely leptonic
channel at LHC even with increased luminosity~
\cite{Birkedal:2004au,Alves:2008up,Asano:2010ii}, while $W'$ can be
observed in the WBF $W^\pm Z \to W^\pm Z$ elastic scattering
~\cite{Birkedal:2004au,Alves:2008up}.  Once a clear signal of the
charged resonance is observed in the above channels, it is mandatory
to study its spin to confirm that the new state is indeed a vector
particle. Much work has been devoted in the literature over the last
years to this issue~\cite{susyspin1,susyspin2,spinreferences}.  For
this purpose the WBF process can be used to determine the spin of a
$W'$ resonance at LHC, however, only for relatively light resonances
and with the assumption of increased luminosity~\cite{Alves:2008up}.
\smallskip

Alternatively the new spin-1 states can also be directly produced in
$pp$ collisions via its coupling to light quarks and in order to
establish that such new vector bosons are indeed associated with EWSB
one should analyze processes in which the new spin-1 decays into
electroweak gauge boson~\cite{Alves:2009aa}.  In this work we
investigate the LHC potential to determine the spin of a new resonance
responsible for the unitarization of weak boson scattering amplitude
by the study of the processes
\begin{equation}
\begin{array}{lll}
pp & \to Z^\prime & \to W^+ W^- \to \ell^+ \ell^{(\prime) -} \,
\Sla{E}_T
\\
pp & \to W^\prime & \to W^\pm Z \to \ell^+ \ell^- \ell^{(\prime) \pm} \,
\Sla{E}_T 
\end{array}\; .
\label{processes}
\end{equation}

Instead of assuming a specific model for EWSB we express our results
as a function of the relevant couplings of the new neutral (charged)
resonance to light quarks and $W^+W^-$ ($W^\pm Z$) pairs, and of its
width and mass. The spin assignment of the new resonances is obtained
from the spin correlation between the final state leptons, contrasting
the expected results for spin--1 and spin--0 new states, {\em i.e.}
we work in the framework commonly used to analyze the spin of
supersymmetric particles~\cite{susyspin1,susyspin2}.  We also study
the angular distribution of the produced EW gauge bosons in the $V'$
center of mass. In order to do so one needs to reconstruct the
neutrino momenta for the processes (\ref{processes}). For the topology
$ \ell^+ \ell^- \ell^{\prime \pm} \, \Sla{E}_T $, associated with the
$W^\prime$ production, the neutrino longitudinal momentum was obtained
requiring that it is compatible with the production of an on-shell
$W$. To reconstruct the momenta of the two neutrinos coming from
$Z^\prime$ production we used the $M_{T2}$ assisted on-shell (MAOS)
reconstruction ~\cite{maos}.  We show that the best sensitivity is
obtained by directly studying the final state leptons and we quantify
the correlated ranges of $V'$ couplings, masses, widths and collider
luminosity for which the spin of the resonance can be established at a
given CL. \smallskip

\section{Framework}

In order to study the processes (\ref{processes}) we must know the
couplings of the new resonance to light quarks and electroweak gauge
bosons, as well as its mass and width. Here we will consider that
these are free parameter without restricting ourselves to any specific
model.  However, for the sake of concreteness we assume that the
couplings of the $Z^\prime$ and the $W^\prime$ to the light quarks and
to gauge bosons have the same Lorentz structure as those of the SM, as
suggested by the Higgsless models, but with arbitrary
strength. Furthermore, we vetoed the $Z^\prime$ coupling to $ZZ$ pairs
as it happens in this class of models. \smallskip

The partial wave amplitude for the process $W^+W^- \rightarrow W^+W^-$
is saturated by the exchange of a $Z^\prime$ provided its coupling to
electroweak gauge bosons satisfies~\cite{Birkedal:2004au}
\begin{equation}
{g_{Z^\prime WW}}_{max}=g_{ZWW}\, \frac{M_Z}{\sqrt{3}M_{Z^\prime}} 
\label{eq:gwwvmax}
\end{equation}
with $g_{ZWW}=g~ c_W$ being the strength of the SM triple gauge boson
coupling.  Here $g$ stands for the $SU(2)_L$ coupling constant and
$c_W$ is the cosine of the weak mixing angle.\smallskip

Analogously, a charged vector resonance saturates unitarity of the
scattering $W^\pm Z \to W^\pm Z$ for~\cite{Birkedal:2004au}
\begin{equation}
{g_{W^\prime WZ}}_{max}
=g_{ZWW}\, \frac{M^2_Z}{\sqrt{3}
M_{W^\prime} M_W}  \; .
\label{eq:gwzvmax}
\end{equation}
In what follows we use ${g_{W^\prime WZ}}_{max}$ and ${g_{Z^\prime
    WW}}_{max}$ simply as convenient normalizations for the coupling
of the spin-1 resonance to SM gauge bosons.\smallskip

The width of the new spin--1 resonances receives contributions from
its decay to light quarks and electroweak gauge boson pairs, as well
as into other states, like $t$ or $b$.  Therefore, in this work we
treat the $Z'$ and $W'$ widths as a free parameters.  In this
approach, for each final state the analysis depends on three
parameters: the mass of the resonance, $M_{V'}$; its width,
$\Gamma_{V'}$; and the product of its couplings to light quarks and to
SM gauge bosons, $g_{V' q\bar q}\; g_{V^{'} W V}$.  These parameters
are only subject to the constraint that for a given value of product
of the couplings of the new resonance and of its mass, there is lower
bound on its width that reads~\cite{Alves:2009aa}
\begin{eqnarray}
&&\Gamma_{Z^\prime}\; >\;0.27\,{\rm GeV}\, \left(
\frac{g_{Z^\prime q\bar q}}{g_{Zq\bar q}}
\right)\,
\left(\frac{g_{Z^\prime WW}}{{g_{Z^\prime WW}}_{max}}\right)\,
\,\left(\frac{M_{Z^\prime}}{M_Z}\right)^2 \; 
\; 
\label{eq:zcouplimit}
\\
&&\Gamma_{W^\prime}\; >\;0.40 \, {\rm GeV}\,\left(
\frac{g_{W^\prime q\bar q}}{g_{Wq\bar q}}
\right)\,
\left(\frac{g_{W^\prime WZ}}{{g_{W^\prime WZ}}_{max}}\right)\,
\,\left(\frac{M_{W^\prime}}{M_W}\right)^2 \;\;\;\;
\label{eq:wcouplimit}
\end{eqnarray} 
where $g_{Zq\bar q}=g/c_W$ and $g_{Wq\bar q^\prime}=g/\sqrt{2}$.  \smallskip

Within our approach we can express the cross section for the processes
(\ref{processes}) as
\begin{eqnarray}
\sigma_{\rm tot}= && \sigma_{SM}\; +\; 
\left(\frac{g_{V^\prime q\bar q}}{g_{Vq\bar q}} \frac{g_{V^\prime WV}}{
{g_{V^\prime WV}}_{max}}\right) 
\, \sigma_{int}(M_{V^\prime},\Gamma_{V^\prime})
\nonumber
\\
&&\;+\;
\left(\frac{g_{V^\prime q\bar q}}{g_{Vq\bar q}} \frac{g_{V^\prime WV}}{
{g_{V^\prime WV}}_{max}}\right)^2
\sigma_{V^\prime}(M_{V^\prime},\Gamma_{V^\prime}) 
\label{eq:sigmatot}
\end{eqnarray}
where the Standard Model, interference and new resonance contributions
are labeled $SM$, $int$ and $V^\prime$ respectively.  Moreover, for
$V^\prime=Z^\prime$, $g_{V^\prime WV}\equiv g_{Z^\prime WW}$ and
$g_{V^\prime q\bar q}\equiv g_{Z^\prime q\bar q}$ while for
$V^\prime=W^\prime$, $g_{V^\prime WV}\equiv g_{W^\prime WZ}$ and
$g_{V^\prime q\bar q}\equiv g_{W^\prime q\bar q^\prime}$.  \smallskip

By construction our analysis applies to any $V^\prime$ whose couplings to
the SM u- and d-quarks and to the SM gauge bosons are a simple rescaling 
of the W or Z couplings. Conversely the analysis renders limited 
information on the underlying physics associated to the new resonances unless
combined with information from  other channels for the observation of 
these states.  For instance, the processes
(\ref{processes}) give information on the couplings to light quark
pairs and electroweak gauge bosons. Analyzing the weak boson fusion
production of these particles allow us to disentangle the couplings to
gauge bosons and quarks. Certainly additional information can be
gathered by studying further channel like the associated production
with a gauge boson or the new resonance decay into leptons.


We perform our analyses at the parton level, keeping the full helicity
structure of the amplitude. This is achieved using the package
MADGRAPH~\cite{madevent} modified to include the new vector states and
their couplings.  In our calculations we use CTEQ6L parton
distribution functions~\cite{CTEQ6} with renormalization and
factorization scales $\mu_F^0 = \mu_R^0 =
\sqrt{({p^{\ell^+}_{T}}^2+{p^{\ell^-}_{T}}^2)/2}$ where
$p^{\ell^\pm}_{T}$ is the transverse momentum of the two charged
leptons in the $Z'$ decay or of the two different flavor opposite sign
leptons in the $W'$ decay. For the case of $W'$ decaying into three
equal flavor leptons we choose the two opposite sign leptons whose
invariant mass is not compatible with being the decay products of a
$Z$.  Furthermore, we simulate experimental resolutions by smearing
the energies, but not directions, of all final state leptons with a
Gaussian error given by a resolution $\Delta E/E = 0.1/\sqrt{E} \oplus
0.01$($E$ in GeV).  We also consider a lepton detection efficiency of
$\epsilon^\ell=0.9$. \smallskip

\section{$W^\prime$ spin determination}

We analyzed $W^\prime$ production in the channel
\[
   pp \to W^\prime \to Z W^\pm \to \ell^+ \ell^- \ell^{\prime \pm}
\Sla{E}_T 
\]
with $\ell= e$ or $\mu$. The main SM backgrounds are the production of
electroweak gauge boson pairs $W^\pm Z$ and $ZZ$ with its subsequent
leptonic decay. In the $ZZ$ production one of the final state leptons
must evade detection. The SM production of top quarks can also lead to
trilepton final states, however, this process is rather suppressed
since one of the isolated leptons must originate from the
semi-leptonic decay of a b quark. \smallskip

The starting cuts are meant to ensure the detection and isolation of
the final leptons plus a minimum transverse momentum:
\begin{equation}
|\eta_\ell| < 2.5 \;,\;
\Delta R_{\ell \ell}> 0.2 \;,\; 
p_{T}^\ell > 10 \hbox{ GeV}\;\hbox{and}\;
\Sla{E}_T > 10 \hbox{ GeV}
\label{basiccuts}
\end{equation}
Next, we look for a same flavor opposite charge lepton pair
that is compatible with a $Z$, {\em i.e.}
\begin{equation}
\left | M_{\ell^+\ell^-}   - M_Z \right | < 20 \hbox{ GeV.}
\label{mllmz}
\end{equation}
We also demand in the search for the resonance that the hardest
observed lepton has transverse momentum in excess of 120 GeV in order
to tame the SM backgrounds.\smallskip

In this process the neutrino momentum can be reconstructed up to a
two--fold ambiguity: its transverse momentum can be directly obtained
from momentum conservation in the transverse directions while its
longitudinal component can be inferred by requiring that
$(p^\nu+p^\ell)^2 = M_W^2$, which leads to
\begin{eqnarray}
p_L^\nu&=&\frac{1}{2 {p^l_T}^2}
\bigg\{\big[M_W^2+2(\vec {p^l_T} \cdot \vec{\Sla{p_T}})\big] p_L^l \nonumber \\
&&\pm 
\sqrt{\big[M_W^2+2(\vec{p_T^l} \cdot \vec{\Sla{p_T}})\big]^2 |\vec p^l|^2 -
4 (p_T^l E^l\Sla{E_T})^2}\bigg\}
\label{eq:plnu}
\end{eqnarray}
where $p^\ell$ is the four-momentum of the charged lepton not
associated to the $Z$. With the two values of the reconstructed
neutrino momenta we obtain two possible solutions for the invariant
mass of the $\ell\ell\ell\nu$ system.  In order to enhance the signal
and reduce the SM backgrounds we require that the final state is
compatible with a $W^\prime$ production of a given mass,
\begin{equation}
  | M^{\rm min}_{\rm rec} - M_{W^\prime}| < \delta \;.
\label{eq:mrec}
\end{equation}
where $ M^{\rm min}_{\rm rec}$ is the smaller of the two solutions.
In our analysis we consider three reference $W^\prime$ masses 500 GeV,
1 TeV, and 1.5 TeV, and we took $\delta = 50$, 100, and 200 GeV for the
three cases, respectively. \smallskip

\begin{figure}[t]
\includegraphics[width=3.5in]{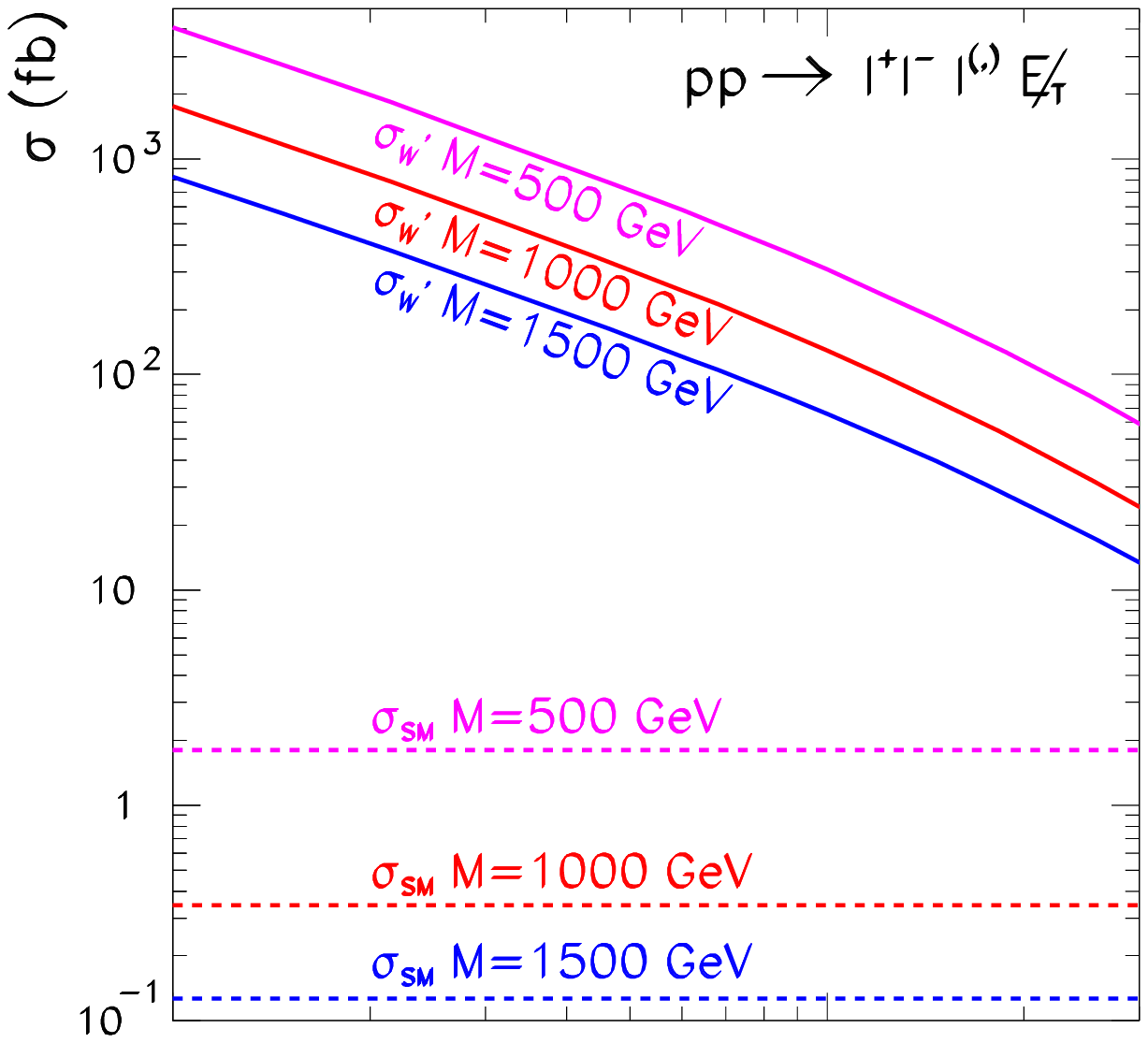}
\vskip -2cm
\includegraphics[width=3.5in]{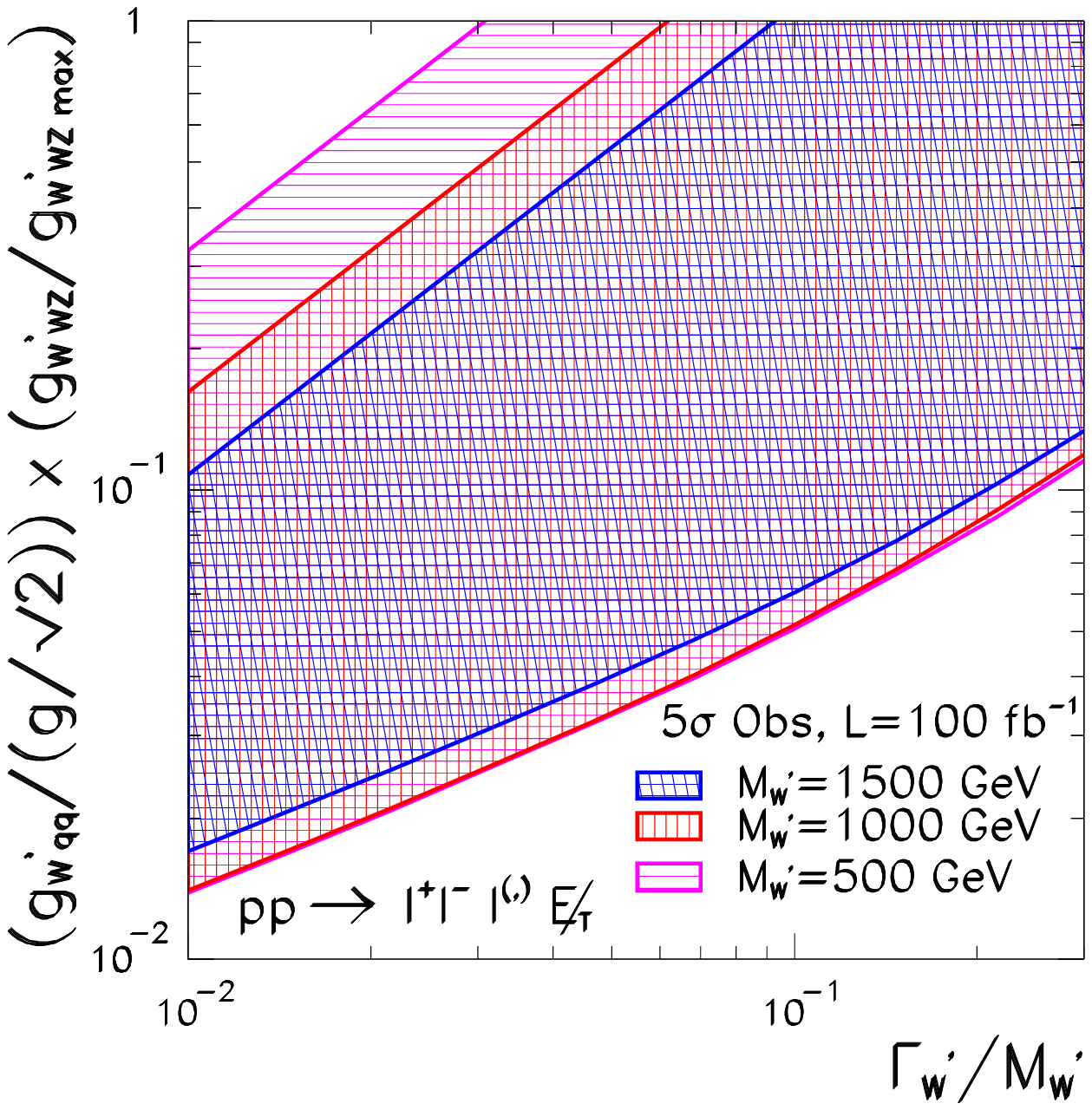}
\caption{{\bf Upper panel:} signal and background cross sections for
  the $\ell^+ \ell^- \ell^{\prime\pm}\,\Sla{E}_T$ final state for all
  possible lepton combinations without including lepton detection
  efficiencies.  {\bf Lower panel}: the filled regions are the ranges
  of the parameters for observation of a $W^\prime$ with mass
  $M_{W^\prime}=0.5$, 1, and 1.5 TeV with at least 5$\sigma$
  significance in the reaction $pp \to W^\prime \to Z W^\pm \to \ell^+
  \ell^- \ell^{\prime\pm}\,\Sla{E}_T$ and an integrated luminosity of
  100 fb$^{-1}$.  }
\label{fig:sensi:wp}
\end{figure}

We show in Fig.~\ref{fig:sensi:wp} (upper panel) the values of
$\sigma_{W'}(M_{W'},\Gamma_{W'})$ and $\sigma_{SM}$ (which after cut
(\ref{eq:mrec}) is also a function of $M_{W'}$) at $\sqrt{s}= 14$ TeV.
Once the cuts described above are imposed the interference term is
negligible for all values of $W'$ mass and widths considered.  As seen
from this figure the SM backgrounds diminish as the new state becomes
heavier, as expected, and the signal cross section deteriorates as the
width of the resonance grows.  Moreover, this channel presents a small
SM background due to the reduced leptonic branching ratio of the
electroweak gauge bosons.  For the sake of completeness we depict in
Fig.~\ref{fig:sensi:wp} (lower panel) the region of the parameter
space where the LHC will be able to observe a $W^\prime$ with at least
$5\sigma$ significance level for an integrated luminosity of 100
fb$^{-1}$.  For this luminosity the number of background events is
large enough for Gaussian statistics to hold for $M_{W'}=500$ and
$1000$ GeV and we impose $N_{W'}\geq 5 \sqrt{N_{\rm SM}}$ where
$N_{W',\rm SM}={\cal L}\times \sigma_{W',\rm SM} \times
(\epsilon^{\ell})^3$.  For $M_{W'}=1500$ GeV the number of expected
background events is $N_{\rm SM}=9.8$ and we adopt the corresponding
5$\sigma$ observability bound for Poisson statistics in the presence
of this background, {\em i.e.}  $N_{W'}>18$. As expected larger
couplings are required for the observation as the resonances
broaden. The upper bounds on the discovery region are due to the
constraint (\ref{eq:wcouplimit}) on the couplings for a given
$W^\prime$ width. As a final remark, with a reduced integrated
luminosity of 10 fb$^{-1}$ the lower line of minimum coupling constant
product needed for discovery is increased by a factor $\simeq 3$,
however, a sizable fraction of the parameter space can still be
probed. \smallskip

It is interesting to compare the results depicted in the lower panel
of Fig.~\ref{fig:sensi:wp} with the direct searches for a $W^\prime$
performed so far.  A comparison of these searches, unfortunately, is
model dependent since the experimental analyses relied on a specific
model.  For instance, they used the direct interactions of the new
$W^\prime $ states with leptons, which is not present in our
parameterization. The only exception is the CDF Collaboration search
for new $WW$ and $WZ$ resonances in $p \bar{p} \to e^\pm j j
\Sla{E}_T$~\cite{cdf}. This work excludes a narrow 500 GeV
$W^\prime$ at 95\% CL provided
\[
\left (
\frac{g_{W^\prime q\bar q}}{g/\sqrt{2}}
\, \times
\frac{g_{W^\prime WZ}}{{g_{W^\prime WZ}}_{max}}\right)\,
 \gsim 0.21   
\]
This implies that a small left corner
of Fig.~\ref{fig:sensi:wp} is probably already excluded.
\smallskip

\begin{figure}[t]
\includegraphics[width=3.5in]{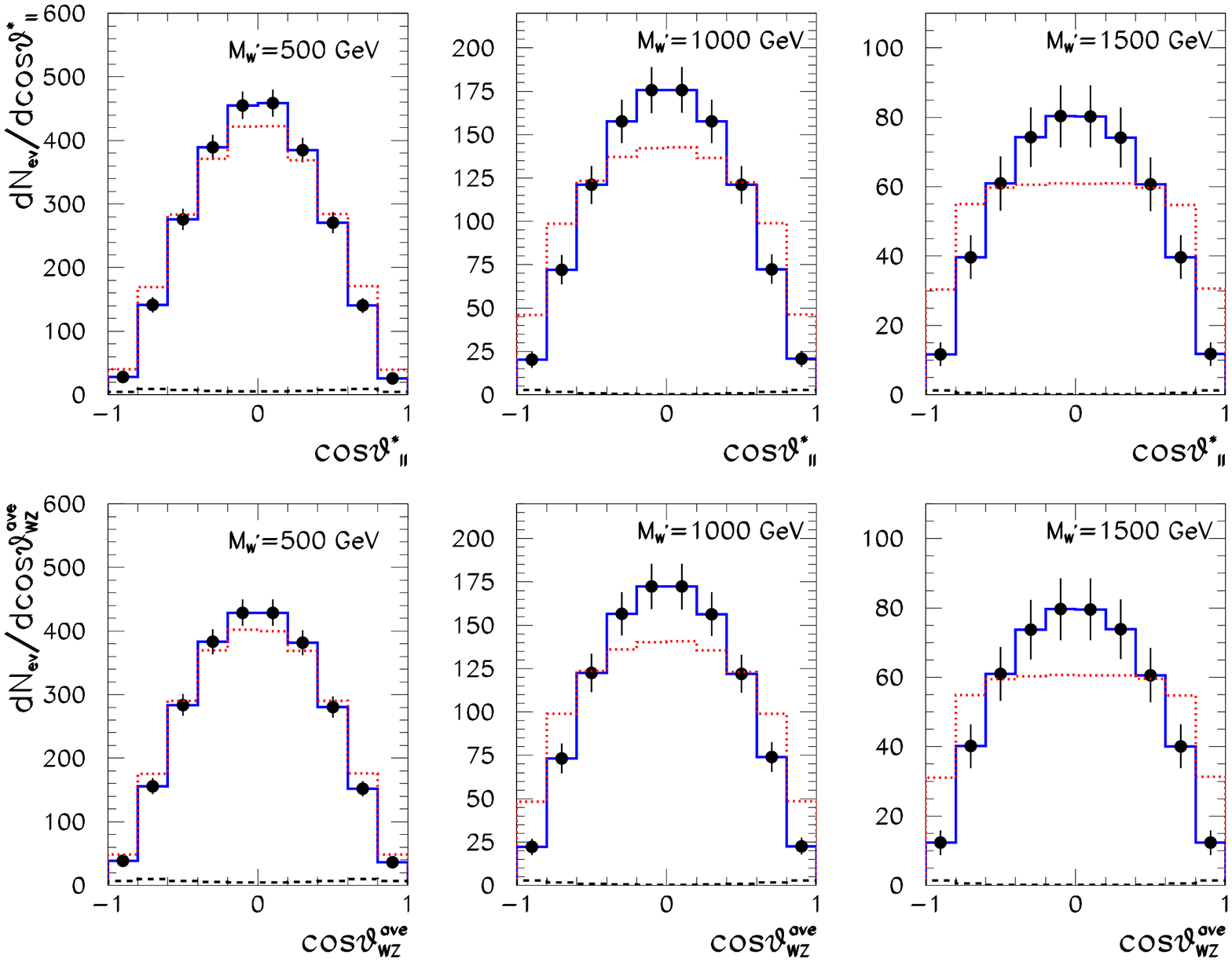}
\caption{$\cos\theta^*_{ll}$ (upper panels) and
  $\cos\theta^{ave}_{WZ}$ (lower panels) distributions for the
  production of the charged vector resonance $W'$ (solid blue line
  with error bars), and the production of a charged scalar resonance
  (dotted red line).  The results are shown for
  $\Gamma_{W^\prime}=0.05 M_{W^\prime}$ and $\left(\frac{g_{W^\prime
        q\bar q'}}{g_{Wq\bar q'}} \frac{g_{W^\prime WZ}}{ {g_{W^\prime
          WZ}}_{max}}\right)=0.3$.  The SM contribution (barely
  visible) is the dashed black line at the bottom. Here we assumed an
  integrated luminosity of 100 fb$^{-1}$. }
\label{fig:ang:wp}
\end{figure}

In previous studies~\cite{susyspin1}, it has been shown that a
convenient variable for contrasting the production of particles with
different spin is
\begin{equation}
   \cos\theta_{\ell\ell}^*\equiv \tanh\left ( \frac{\Delta\eta_{\ell\ell}}{2}
   \right)\; ,
\label{eq:barr}
\end{equation}
where $\Delta\eta_{\ell\ell}$ is the rapidity difference between the
same charge leptons. This quantity has the advantage of being
invariant under longitudinal boosts.  We present in the upper panels
of Fig.~\ref{fig:ang:wp} the $\cos\theta_{\ell\ell}^*$ spectrum for
the production of spin--0 and spin--1 resonances and our three
reference masses. 
In order to compare the spin--0 and spin--1 angular
correlations we assumed that the production cross section of the
spin--0 particles is equal to the one for spin--1 particles, and
we imposed that the scalar and vector resonances have the same mass
and width. 
As we can see, the $\cos\theta_{\ell\ell}^*$ distribution
for $W^\prime$ vector production exhibits a maximum at
$\cos\theta_{\ell\ell}^*=0$, as expected. In principle this spectrum
should be flat in the production of scalars, however, the acceptance
cuts, especially $|\eta_\ell|< 2.5$, distort this spectrum, which
reduces the discriminating power for light resonances. \smallskip

\begin{figure}[t]
\includegraphics[width=3.5in]{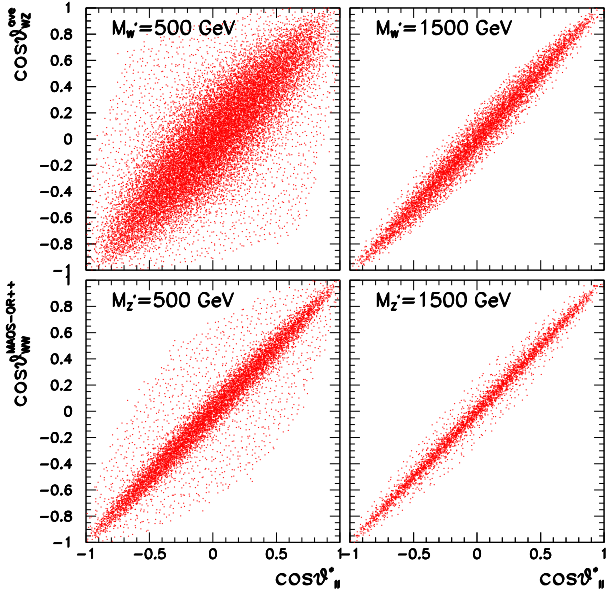}
\caption{The upper panels contain the $\cos\theta^*_{ll} \otimes
  \cos\theta^{ave}_{WZ}$ spectrum for $W^\prime$ and
  $M_{W^\prime}=0.5$ TeV (upper left panel) and 1.5 TeV (upper right panel) where
  $\cos\theta^{ave}_{WZ}$ is the average of the two possible
  solutions. The lower panels depict the $\cos\theta^*_{ll} \otimes
  \cos\theta^{MAOS-OR++}_{WW}$ spectrum for $Z^\prime$ and
  $M_{Z^\prime}=0.5$ TeV (lower left panel) and 1.5 TeV (lower right panel).}
\label{fig:scat}
\end{figure}

The extraction of the final state neutrino momentum allow us to
reconstruct angular correlations in the $WZ$ center--of--mass
frame. Therefore, we also study the spin correlations using the
reconstructed $Z$ polar angle ($\theta_{WZ}$) distribution evaluated
in the $WZ$ center--of--mass frame.  Since there is a two--fold
ambiguity in this reconstruction, we consider the average of the two
resulting distributions in our analysis. As shown in
Ref.~\cite{Alves:2008up} the angular distribution of the reconstructed
$\cos\theta_{WZ}$ for the reconstruction yielding minimum (maximum)
$WZ$ invariant mass, is peaked (has a valley) around zero when
compared to the true $\theta_{WZ}$ but the average of the two has a
very similar distribution to the true one.  We plot in the lower
panels of Fig.~\ref{fig:ang:wp} the $\cos\theta_{WZ}^{ave}$ spectrum
for the production of spin--0 and spin--1 resonances and our three
reference masses. Comparing the distributions in the two angular
variables $\cos\theta^*_{ll}$ and $\cos\theta_{WZ}^{ave}$ we learn
that they are very similar.  Indeed they happen to be strongly
correlated as shown in the upper panels of Fig.~\ref{fig:scat}, where we
plot the $\cos\theta^*_{ll} \otimes \cos\theta_{WZ}^{ave}$ spectrum
for $M_{W^\prime} = 0.5$ TeV (upper left panel) and 1.5 TeV (upper right
panel). The figure is for $\Gamma_{W'}=0.05 M_{W'}$ but the results
are very insensitive to the precise value of the width.  It is clear
from Fig.~\ref{fig:scat} that there is a strong correlation between
$\cos\theta^*_{\ell\ell}$ and $\cos\theta_{WZ}^{ave}$, which is
somehow unforeseen given the definitions of both variables and the
different behaviours of the $\cos\theta_{WZ}^{max}$ and
$\cos\theta_{WZ}^{min}$ distributions.  As expected the correlation
gets stronger as the $W^\prime$ mass increases since heavier
resonances decay into more energetic electroweak gauge bosons and
consequently the final state leptons have the tendency to follow the
direction of the parent $W$ or $Z$.  \smallskip

Taking into account the correlation between $\cos\theta^*_{ll}$ and
$\cos\theta^{ave}_{WZ}$, it is expected that both kinematical
variables have a comparable spin discriminating power. In fact, this
is the case except for $M_{W'}=500$ GeV, where $\cos\theta^{ave}_{WZ}$
performs slightly worse.  Furthermore we should keep in mind that
larger systematic uncertainties are expected in the reconstruction of
$\theta_{WZ}$ associated with the understanding and calibration of the
detector due to the measurement of missing transverse
momentum. \smallskip

In order to quantify the parameter space region for which a positive
discrimination between spin--0 and
spin--1 resonances is possible we
construct the asymmetry
\begin{equation}
A_{\ell\ell} = 
\frac{\sigma(|\cos\theta^*_{\ell\ell}| < 0.5) - \sigma(|\cos\theta^*_{\ell\ell}| 
> 0.5) }
{\sigma(|\cos\theta^*_{\ell\ell}| < 0.5) + \sigma(|\cos\theta^*_{\ell\ell}| 
> 0.5) } \; .
\label{asym:emu}
\end{equation}
Notice that this observable eliminates possible normalization
systematics in the angular distributions. \smallskip

\begin{figure}[t]
\includegraphics[width=3.5in]{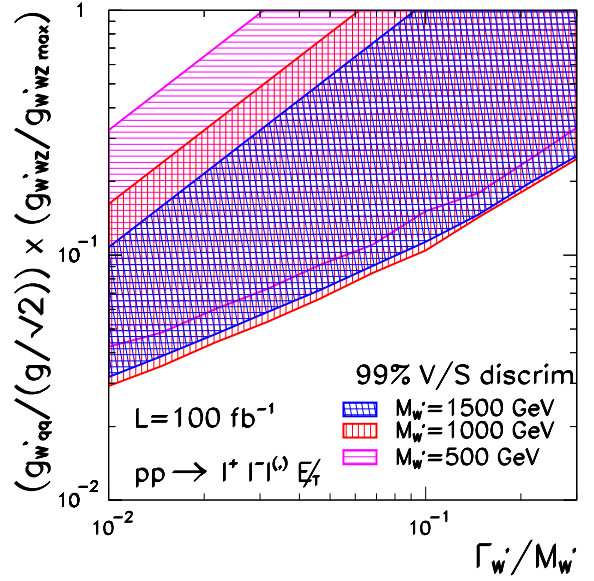}
\caption{Parameter space region where the $W^\prime$ spin can be
  determined with 99\% CL using the asymmetry $A_{\ell\ell}$ for an
  integrated luminosity of 100 fb$^{-1}$.  }
\label{fig:discwp:ll}
\end{figure}

Fig.~\ref{fig:discwp:ll} displays the region in the parameter space
where the $W^\prime$ spin can be established with 99\% CL using
$A_{\ell\ell}$ for an integrated luminosity of ${\cal L}$=100
fb$^{-1}$ at the LHC. This result was obtained taking into account
only the statistical errors and assuming that the observed
distribution follows that of a vector resonance. With this hypothesis
the 99\% CL spin discrimination condition reads:
\begin{equation}
  |A^V_{\ell\ell}- A^S_{\ell\ell}| \geq 2.58 ~\sigma_{A^V_{\ell\ell}}=
  2.58~ \frac{\sqrt{1-{A^V_{\ell\ell}}^2}}{\sqrt{N_{\rm tot}}}
\label{eq:spindis}
\end{equation}
where $\sigma_{A^V_{\ell\ell}}$ is the expected statistical error of
the variable $A^V_{\ell\ell}$ and $N_{\rm tot}={\cal L}\times
\sigma_{\rm tot} \times (\epsilon^\ell)^3$ with $\sigma_{\rm tot}$ in
Eq.~(\ref{eq:sigmatot}).  In writing Eq.~(\ref{eq:spindis}) we
implicitly assume that for the 99\% spin determination the number of
events $N_{\rm tot}$ is always large enough for Gaussian statistics to
hold. We verify that this is the case even for the smallest couplings
for which 99\% CL spin determination is possible and therefore the
procedure is consistent. \smallskip

Comparing Figs.~\ref{fig:discwp:ll} and \ref{fig:sensi:wp} we see that
the minimum couplings necessary to determine the spin at 99\% CL for
this integrated luminosity is a factor of an order of 2 larger than
the minimum couplings needed for its discovery. Moreover, as seen in
Fig.~\ref{fig:ang:wp}, the acceptance cuts modify more drastically the
distributions for lighter $W^\prime$ masses, and consequently, the
discrimination between spin--0 and spin--1 requires a larger
statistics, reflected in larger couplings and production cross
sections. Notwithstanding, the LHC will be able to successfully
unravel the spin of a possible new state with 99\% CL in a large
fraction of the parameter space of discovery. \smallskip

\begin{figure}[t]
\includegraphics[width=3.5in]{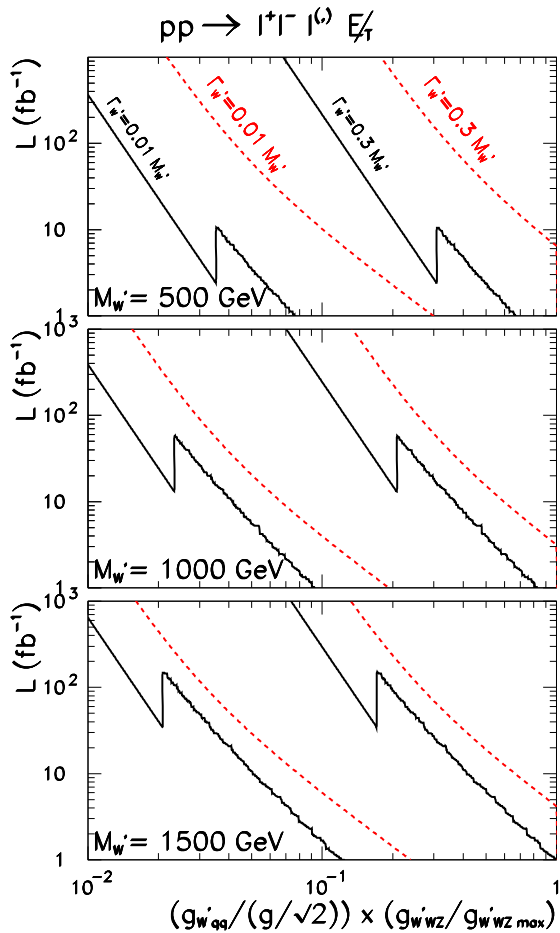}
\caption{The solid (dashed) lines stand for integrated luminosity
  required for the discovery (99\% CL spin determination) as a function
  of the vector resonance couplings. We present the results for three
  masses and two widths: $\Gamma_{W'}=0.01 M_{W^\prime}$ and
  $\Gamma_{W'}=0.3 M_{W^\prime}$. See text for detailed information
  on the statistics used in this figure.}
\label{fig:lum:wp}
\end{figure}

In order to address the potential of the LHC from earlier runs or with
upgraded luminosity, but still at 14 TeV, we quantify the luminosity
requirement for discovery and spin determination of the resonance as a
function of its parameters.  Fig.~\ref{fig:lum:wp} depicts the
integrated luminosity needed for a 5$\sigma$ discovery (solid lines)
and 99\% CL spin determination based on (\ref{eq:spindis}) (dashed
lines) as a function of its couplings for our three reference masses
and two widths ( $\Gamma_{W'}=0.01 M_{W^\prime}$ and $\Gamma_{W'}=0.3
M_{W^\prime}$).  The discovery requirements were obtained using
Poisson or Gaussian statistics depending on whether the expected
number of SM events was smaller or larger than 15 and the changing
from one to the other determines the discontinuity in the
corresponding lines.  For the 99\% CL spin determination the number of
expected events is always large enough for Gaussian statistics to
hold. As we can see from Fig.~\ref{fig:lum:wp}, an earlier discovery, {\em e.g.}
with 10 fb$^{-1}$, is possible even for rather weakly coupled
$W^\prime$.  Although the $W^\prime$ spin determination requires
larger couplings it can also be carried out in a sizable region of the
parameter space in earlier runs. \smallskip

\section{$Z^\prime$ spin determination}

We study the $Z^\prime$ spin through the reaction
\[
   pp \to Z^\prime \to W^+ W^- \to \ell^+ \ell^{(\prime)-} \Sla{E}_T  \; .
\]
The main SM backgrounds to this process are the production of $W^+W^-$
with its subsequent leptonic decay, the $ZZ$ production with one Z
decaying into charged leptons and the other decaying invisibly or with
both Z decaying into charged leptons two of which escape undetected.
Additional backgrounds are provided by the SM production of $t\bar{t}$
pairs with both top quarks decaying semileptonically as well as the
$\tau^+\tau^-$ production with both $\tau$'s decaying
leptonically.\smallskip

We begin our analysis requiring two final state leptons with opposite
charge and applying acceptance and isolation cuts on them
\begin{equation}
|\eta_\ell| < 2.5 \;\;,\;\;
\Delta R_{\ell \ell}> 0.2\;\;\;\hbox{and}\;\; 
p_{T}^\ell > 50 \hbox{ GeV}
\label{cutsww1}
\end{equation}

The presence of two neutrinos in the final state renders impossible
the complete reconstruction of the event. A possible variable to
characterize the signal is the transverse invariant mass,

\begin{eqnarray}
  M_T^{WW} 
=&& \biggl[ \left( \sqrt{(p_T^{\ell^+\ell^{\prime -}})^2 + 
m^2_{\ell^+ \ell^{\prime -}}}   
              + \sqrt{\Sla{p_T}^2 + m^2_{\ell^+\ell^{\prime -} }} \right)^2 \biggr .
\nonumber \\ 
&&\biggl . - (\vec{p}_T^{~\ell^+\ell^{\prime -}} + 
\vec{\Sla{p_T}}  )^2 \biggr]^{1/2} 
\label{eq:mtww}
\end{eqnarray}
where $\vec{\Sla{p_T}}$ is the missing transverse momentum vector,
$\vec{p}_T^{~\ell^+\ell^{\prime -}}$ is the transverse momentum of the
pair $\ell^+ \ell^{\prime -}$ and $m_{\ell^+\ell^{\prime -}}$ is the
$\ell^+ \ell^{\prime -}$ invariant mass. \smallskip

Alternatively we attempt to reconstruct the $WW$ invariant mass 
by estimating the momenta of the two escaping neutrinos
produced using the $M_{T2}$ Assisted on-Shell
(MAOS)  reconstruction~\cite{maos}. For $W^+ (p_1+p_2)
W^- (k_1+k_2) \to \ell^+(p_1)
\nu(p_2) \ell^-(k_1) \nu(k_2)$ the variable $M_{T2}$ is defined as
~\cite{Lester:1999tx}
\begin{equation}
M_{T2} 
\equiv \begin{array}{c}
\min\limits_{{\mathbf{p_2}_T+\mathbf{k_2}_T} =
\Sla{{\mathbf{p}}_T}} \end{array}
\left [{\rm max }
\left \{
M_T( \mathbf{p_1}_T,\mathbf{p_2}_T), 
M_T( \mathbf{k_1}_T,\mathbf{k_2}_T)\right \} \right ]
\end{equation}
where $M_T$ is the transverse mass
\begin{equation}
  M_T^2( \mathbf{p_1}_T,\mathbf{p_2}_T) = 2 (
  | \mathbf{p_1}_T |  | \mathbf{p_2}_T | -
   \mathbf{p_1}_T \cdot \mathbf{p_2}_T ) \;.
\end{equation}
For an event without initial state radiation the transverse MAOS
momenta are simply given by
\begin{equation}
{\bf p_2}^{\rm maos}_T = - {\bf k_1}_T,\quad {\bf
k_2}^{\rm maos}_T = - {\bf p_1}_T.
\label{eq:maost} 
\end{equation}

There can be two different schemes to define the longitudinal MAOS
momenta.  One is to require the on-shell conditions for both the
invisible particles in the final state and the mother particles in the
intermediate state ($W$) ~\cite{maos} (here called MAOS-original )
which results into a four-fold degeneracy
\begin{eqnarray}
&& {p_2}^{\rm maos}_L(\pm) = \frac{1}{|{\bf
p_1}_T|^2}\left[ p_{1L}\,A \right. \nonumber \\
&& \left.\pm \sqrt{|{\bf
p_1}_T|^2+p_{1L}^2}\,\sqrt{A^2-|{\bf p_1}_T|^2|{\bf p_2}^{\rm maos}_T|^2}
\right]\,, \nonumber \\
&& {k_2}^{\rm maos}_L(\pm) = \frac{1}{|{\bf k_1}_T|^2}
\left[ k_{1L}\,B \right. \nonumber \\
&& \left.
\pm
\sqrt{|{\bf k_1}_T|^2+k_{1L}^2}\,\sqrt{B^2-|{\bf k_1}_T|^2|{\bf k_2}^{\rm
maos}_T|^2} \right]\,,
\label{eq:maoslo} 
\end{eqnarray}
where $A\equiv M_W^2/2+{\bf p_1}_T\cdot{\bf p_2}^{\rm maos}_T$ and
$B\equiv M_W^2/2+{\bf k_1}_T\cdot{\bf k_2}^{\rm maos}_T$. \smallskip

Another possible scheme~\cite{maos2}  is to require
\begin{eqnarray}
&& (p_{2}^{\rm maos})^2=(k_{2}^{\rm maos})^2=0 \; ,  \\ \nonumber 
&& (p_1+p_{2}^{\rm
maos})^2=(k_1+k_{2}^{\rm maos})^2=M_{T2}^2,
\end{eqnarray}  
which gives unique longitudinal MAOS momenta  
(here called MAOS-modified) as
\begin{equation}
{p_2}^{\rm maos}_L=\frac{|{\bf p_2}^{\rm
maos}_T|}{|{\bf p_1}_T|}p_{1L},\quad {k_2}_L^{\rm maos}=\frac{|{\bf k_2}^{\rm
maos}_T|}{|{\bf k_1}_T|}k_{1L}.
\label{eq:maoslm} 
\end{equation}

\begin{figure}[t]
\includegraphics[width=3.5in]{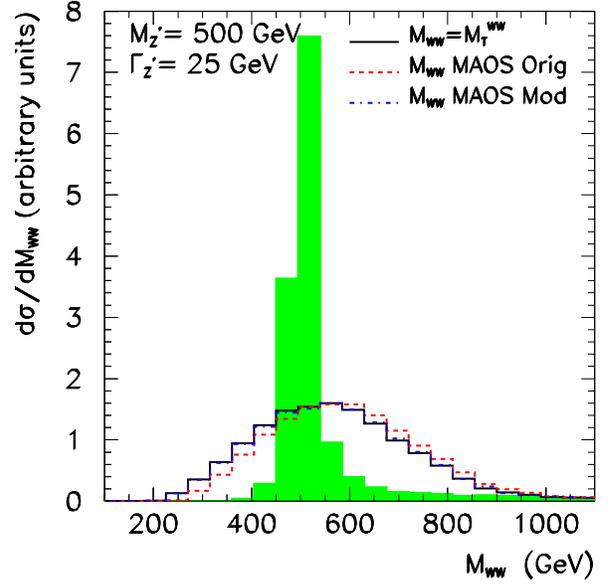}
\caption{Reconstructed $WW$ invariant mass distributions for $pp \to
  Z^\prime \to W^+ W^- \to \ell^- \ell^+ \Sla{E}_T$ assuming
  $M_{Z'}=500$ GeV and $\Gamma_{Z'}=25$ GeV. The solid (black) line
  corresponds to $M_{WW}=M_T^{WW}$ (\ref{eq:mtww}). The dashed (red)
  line stands for the invariant mass reconstructed using the
  MAOS-original momentum prescription with sign ++ in
  Eq.~(\ref{eq:maoslo}) and the dash-dot (blue) line represents
  MAOS-modified (\ref{eq:maoslm}) prescription. The shadow (green)
  area represents the true spectrum. }
\label{fig:mzp}
\end{figure}

To illustrate the accuracy of the neutrino momenta determination in
the MAOS reconstruction we present in Fig.~\ref{fig:mzp} the
reconstructed $Z^\prime\to W^+ W^-$ invariant mass using the
MAOS-original (with sign ++ in Eq.~(\ref{eq:maoslo}) for
illustration), MAOS-modified (\ref{eq:maoslm}), and the $WW$
transverse invariant mass (\ref{eq:mtww}).  For the sake of comparison
the shaded (green) area represents the actual spectrum.  As we can
see, the three methods lead to similar results which is expected since
the signal is dominated by $Z'$ decaying into on-shell
$W$'s. Furthermore, in all cases the characteristic peak associated
with the production of the resonance is substantially broadened.
However it is still possible to suppress the backgrounds and enhance
the $Z^\prime$ signal by demanding that any of the reconstructed $WW$
masses to be around $M_{Z'}$ within a broad width.  \smallskip

Consequently in our study we demand the $WW$ transverse invariant mass to 
comply with
\begin{equation}
  M_T^{WW} 
>\frac{M_{Z^\prime}}{2}  \;,
\label{eq:mtcut}
\end{equation}
where only a lower cut is required because the background is a very steeply
falling function of $M_T^{WW}$.  

After the cuts (\ref{cutsww1}) and (\ref{eq:mtcut}), the $t \bar{t}$
SM background is still quite large, therefore we veto the presence
of additional jets in the event with
\begin{equation}
    | \eta_j | < 3   \;\;\;\; \hbox{ and } \;\;\; p_T^j > 20 \;\hbox{ GeV.}
\label{eq:veto}
\end{equation}
However, QCD radiation and pile-up can lead to the appearance of an
additional jet even in signal events. Therefore, we must introduce the
probability of a QCD (electroweak) event to survive such a central jet
veto~\cite{rainth}.  The survival probability due to pile--up has been
estimated to be 0.75 for a threshold cut of $p_T=20$ GeV in
Ref.~\cite{atlas}.  Taking into account these two effects we included
in our analysis veto survival probabilities
\begin{equation}
P_{\rm surv}^{\rm EW}= 0.56 \;\;\;\;\;\; , \;\;\;\;\;\;
P_{\rm surv}^{\rm QCD}=0.23 \;\; .
\label{eq:psurv}
\end{equation}

For events presenting same flavor lepton pairs, {\em i.e.} $ee$ or
$\mu\mu$, there is an additional SM contribution stemming from $ZZ$
production with one of the $Z$ decaying invisibly and the other into a
charged lepton pair. For these final states, we supplement the cuts
(\ref{cutsww1}), (\ref{eq:mtcut}), and (\ref{eq:veto}) further
imposing that
\begin{equation}
     \Sla{E}_T > 50 \; \hbox{ GeV} \;\;\hbox{and }\;\;
      m_{\ell^+ \ell^-} > 100 \; \hbox{ GeV.}
\label{etcut}
\end{equation}
We denote the sum of the SM backgrounds not originating from $t\bar t$
production as EW background. \smallskip

\begin{figure}[t]
\includegraphics[width=3.5in]{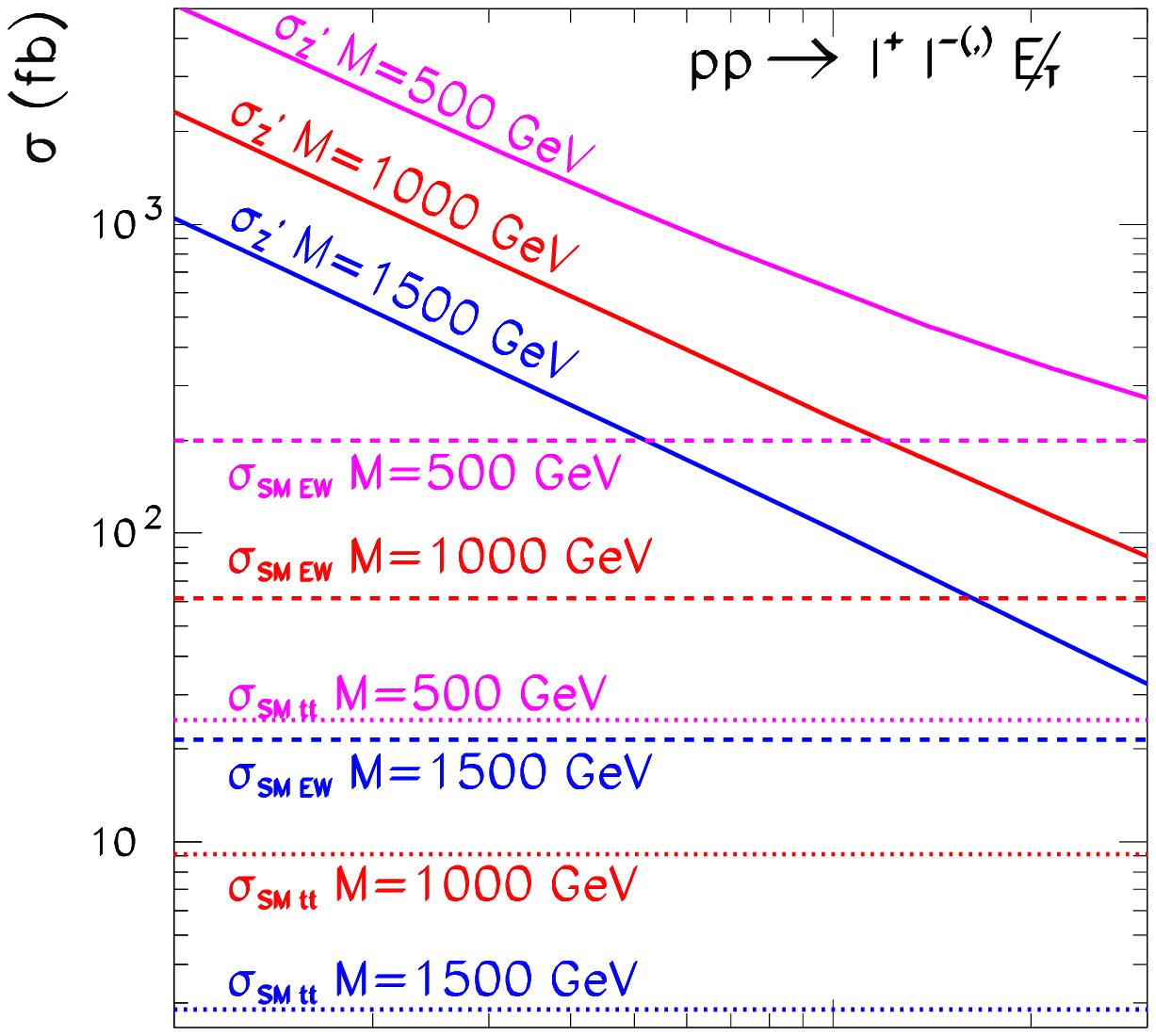}
\vskip -2cm
\includegraphics[width=3.5in]{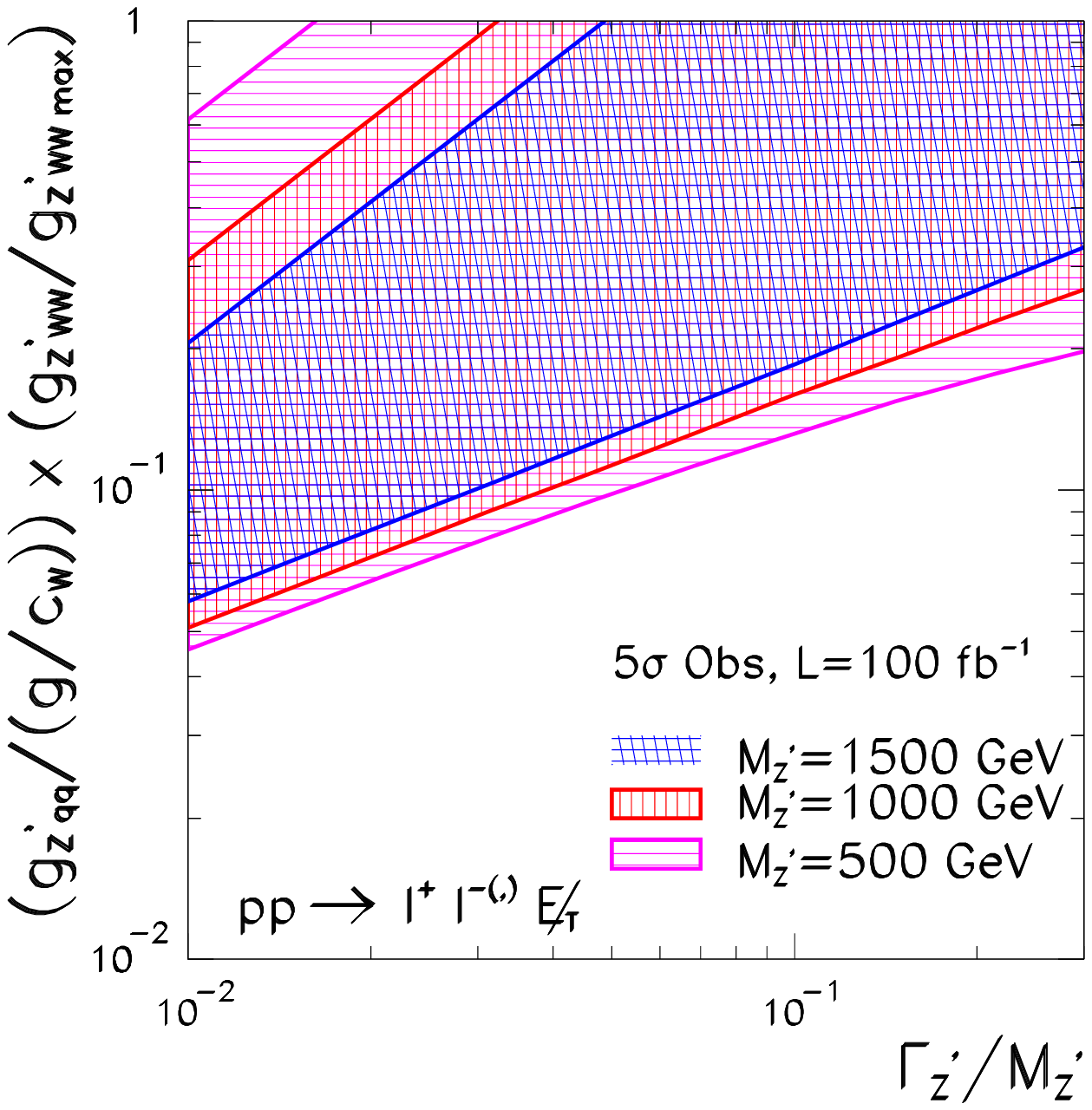}
\caption{{\bf Upper panel:} signal and background cross sections for
  the $\ell^+ \ell^{\prime-}\,\Sla{E}_T$ final state for all possible
  lepton flavor combinations without including lepton detection
  efficiencies nor survival probabilities.  {\bf Lower panel}: the
  filled regions are the ranges of the parameters for observation of a
  $W^\prime$ with mass $M_{W^\prime}=0.5$, 1, and 1.5 TeV with at
  least 5$\sigma$ significance in the reaction $pp \to Z^\prime \to
  W^+W^- \to \ell^+ \ell^{\prime-}\,\Sla{E}_T$ for an integrated
  luminosity of 100 fb$^{-1}$.  }
\label{fig:sensi:zp}
\end{figure}

We show in the upper panel of Fig.~\ref{fig:sensi:zp} the values of
$\sigma_{Z'}$ and $\sigma_{SM}$ for the electroweak and $t\bar t$
backgrounds at $\sqrt{s}= 14$ TeV.  Once the cuts described above are
imposed the interference term is negligible for all values of $Z'$
masses and widths considered.  We see that the backgrounds for $Z'$ in
the leptonic final states are considerably larger than the ones for
$W'$ as a consequence of the very broad reconstruction of the $Z'$
invariant mass in this channel.  For the sake of completeness we
depict in the lower panel of Fig.~\ref{fig:sensi:zp} the parameter
space region where the LHC will be able to observe a $Z^\prime$ with
at least $5\sigma$ significance level for an integrated luminosity of
100 fb$^{-1}$. For this luminosity the number of background events is
always large enough for Gaussian statistics to hold and we impose
$N_{Z'}\geq 5 \sqrt{N_{\rm SM}}$, where $N_{Z'}={\cal L}\times
\sigma_{Z'} \times P_{\rm surv}^{\rm EW} \times (\epsilon^{\ell})^2 $
and $N_{SM}={\cal L}\times \left( \sigma^{\rm EW}_{\rm SM}\times
  P_{\rm surv}^{\rm EW} +\sigma^{t\bar t}_{\rm SM}\times P_{\rm
    surv}^{\rm QCD} \right) \times (\epsilon^{\ell})^2$.  Comparing
with Fig.~\ref{fig:sensi:wp} we find that establishing the existence
of a $Z^\prime$ requires larger couplings to light quark and vector
boson pairs than a $W^\prime$ as a consequence of the larger SM
backgrounds. \smallskip


Nowadays, a small part of the lower panel of Fig.~\ref{fig:sensi:zp}
has been directly probed. The CDF analysis~\cite{cdf} indicates that 
for narrow 500 GeV $Z^\prime$s 
\[
\left (
\frac{g_{Z^\prime q\bar q}}{g_{Zq\bar q}}
\times
\frac{g_{Z^\prime WW}}{{g_{Z^\prime WW}}_{max}}\right)\,
 \gsim 0.19
\]
is excluded at 95\% CL. \smallskip


\begin{figure}[t]
\includegraphics[width=3.5in]{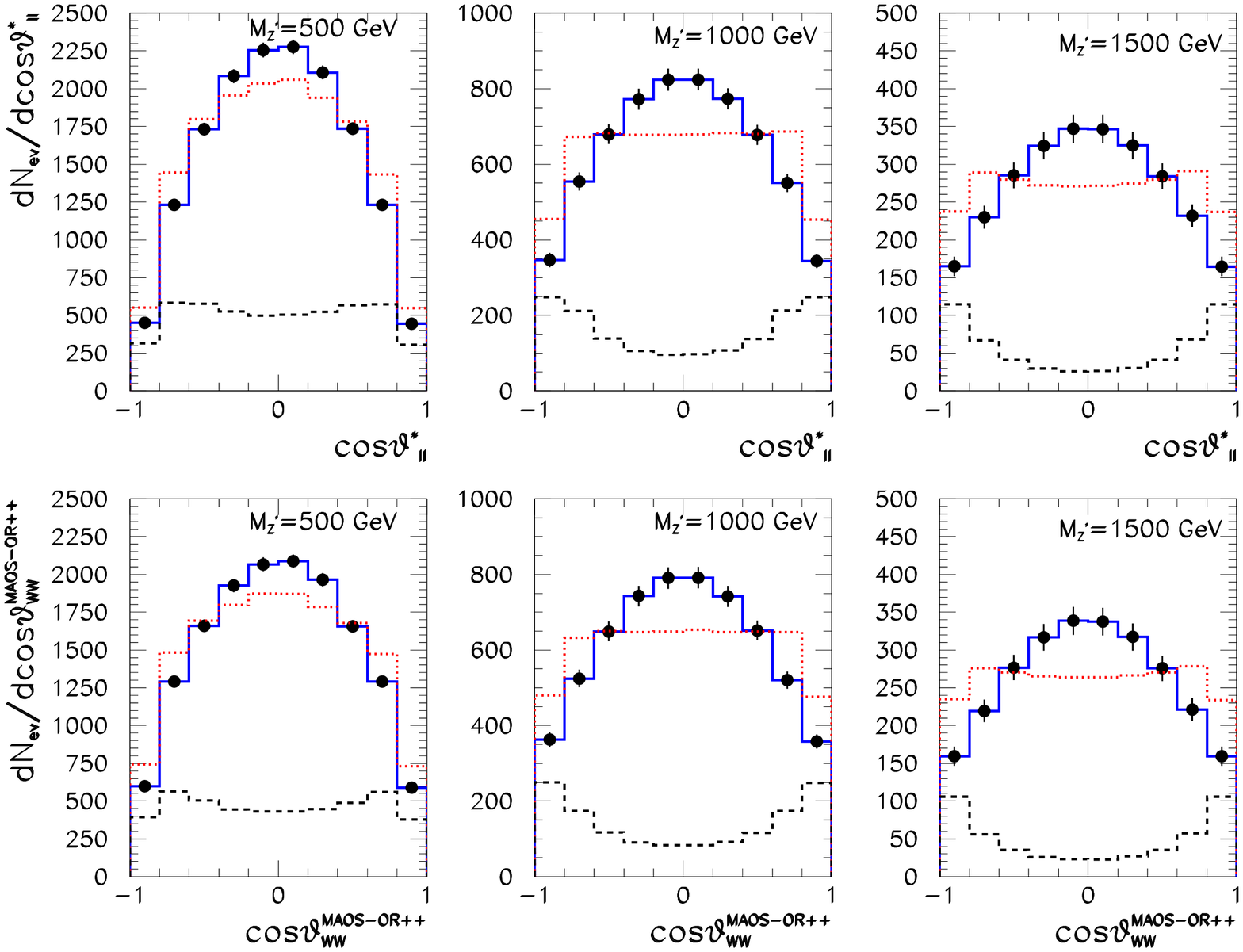}
\caption{$\cos\theta^*_{ll}$ (upper panels) and $\cos\theta_{WW}^{\rm
    MAOS-OR++}$ (lower panels) distributions for the production of the
  neutral vector resonance (solid blue line with error bars), and the
  production of a neutral scalar resonance (dotted red line).  The
  results are shown for $\Gamma_{Z^\prime}=0.01 M_{Z^\prime}$ and
  $\left(\frac{g_{Z^\prime q\bar q}}{g_{Zq\bar q}} \frac{g_{Z^\prime
        WW}}{ {g_{Z^\prime WW}}_{max}}\right)=0.3$.  The contribution
  of the SM background is depicted by the dashed black line. We assumed
  an integrated luminosity of 100 fb$^{-1}$. }
\label{fig:ang:zp}
\end{figure}

In order to discriminate the spin of the neutral resonance we first
employ the variable $\cos\theta_{\ell\ell}^*$ (\ref{eq:barr}) using
the two opposite charge leptons which does not require the
determination of the neutrino momenta, therefore avoiding
reconstruction ambiguities.  We plot in Fig.~\ref{fig:ang:zp} (upper
panels) the $\cos\theta_{\ell\ell}^*$ spectrum for the production of
spin--0 and spin--1 resonances for our three reference masses and
assuming a width of $\Gamma_{Z'}=0.01 M_{Z'}$ and
$\left(\frac{g_{Z^\prime q\bar q}}{g_{Zq\bar q}} \frac{g_{Z^\prime
      WW}}{ {g_{Z^\prime WW}}_{max}}\right)=0.3$. We imposed that the
spin--0 production cross section has the spin--1 value.  Analogously
to the $W^\prime$ case, we can see that the acceptance cuts distort
considerably the spin--0 spectrum at lower masses. For heavier states
the final state leptons have a larger tendency to follow the direction
of the parent $W$ since it is more energetic, ameliorating the effect
of the cuts. Another important feature of this case is that the SM
background is no longer negligible. \smallskip

We have also explored the expected distribution of the $W$ polar angle
in the $W^+W^-$ center--of--mass frame as reconstructed using the
different MAOS prescriptions.  As an illustration we depict, in the
lower panels of Fig.~\ref{fig:ang:zp}, the reconstructed
$\cos\theta_{WW}$ spectrum for the production of spin--0 and spin--1
resonances and our three reference masses as obtained from the
MAOS-original momentum prescription with sign ++ in
Eq.~(\ref{eq:maoslo}).  The comparison of the distributions of the two
angular variables presented in this figure indicates that these
variables are correlated. Indeed they are strongly correlated as can
be seen in the lower panels of Fig.~\ref{fig:scat}.  Consequently, as
in the $W^\prime$ case, we can foresee a similar spin discriminating
power for both variables on the basis of statistics.  We have verified
that the same conclusion is reached when using either the
MAOS-original momentum prescription with sign $--$ in
Eq.~(\ref{eq:maoslo}), the average of the distributions with $+-$ and
$-+$ signs, or the MAOS-modified prescription
(\ref{eq:maoslm}). \smallskip

\begin{figure}[t]
\includegraphics[width=3.5in]{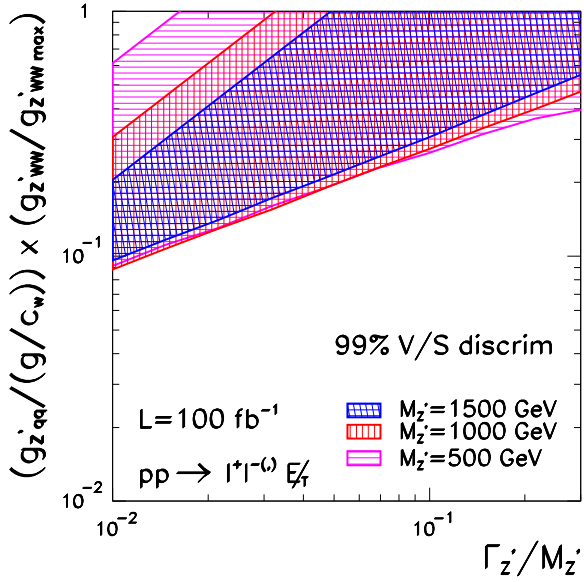}
\caption{Parameter space region where the $Z^\prime$ spin
  determination can be done at 99\% CL for an integrated luminosity
  for 100 fb$^{-1}$ using the asymmetry $A_{\ell\ell}$.}
\label{fig:disczp:ll}
\end{figure}

We present in Fig.~\ref{fig:disczp:ll} the $Z^\prime$ parameter space
region where the LHC can establish its spin with 99\% CL using
$A_{\ell\ell}$ for an integrated luminosity of 100 fb$^{-1}$.  As in
the $W^\prime$ case, the minimum couplings needed for the spin
determination are approximated twice the ones required for the
$Z^\prime$ discovery. Moreover, the minimum couplings required for the
spin determination exhibit a very mild dependence on the resonance
mass since the acceptance cut effects are smaller for heavier states,
compensating, partially, the decrease in the production cross
section. \smallskip

\begin{figure}[t]
\includegraphics[width=3.5in]{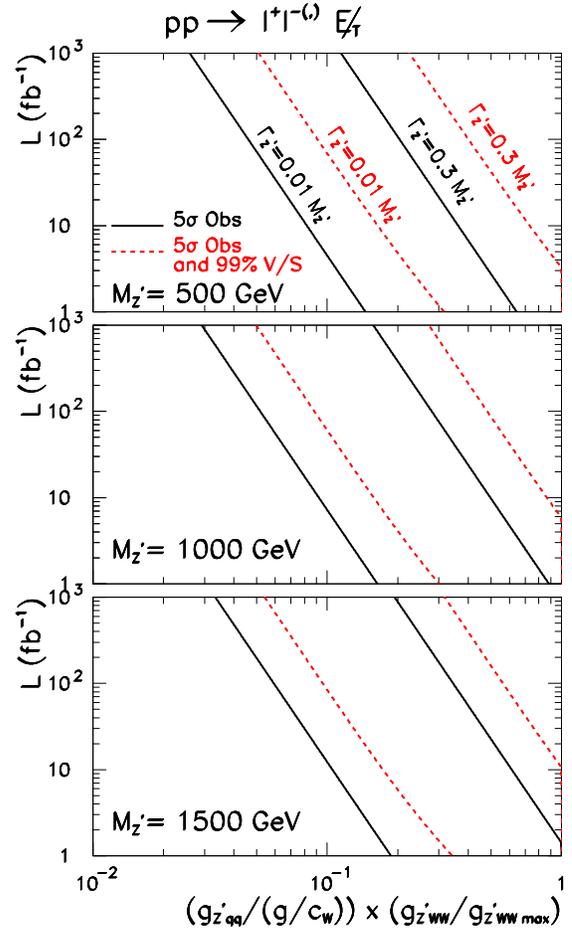}
\caption{The solid (dashed) lines stand for integrated luminosity
  required for the $5\sigma$ discovery (99\% CL spin determination) as
  a function of the vector resonance couplings. We present the results
  for three masses and two widths: $\Gamma_{Z'}=0.01 M_{Z^\prime}$ and
  $\Gamma_{Z'}=0.3 M_{Z^\prime}$.}
\label{fig:lum:zp}
\end{figure}

Finally we show in Fig.~\ref{fig:lum:zp} the required integrated
luminosity for a 5$\sigma$ discovery (solid lines) and 99\% CL spin
determination based on (\ref{eq:spindis}) (dashed lines) for our three
reference masses and two widths ( $\Gamma_{Z'}=0.01 M_{Z^\prime}$ and
$\Gamma_{Z'}=0.3 M_{Z^\prime}$) as a function of the $Z'$
couplings. We find that for a given value of the $Z'$ couplings the
required luminosity for 99\% CL spin determination based on the study
of $A_{\ell\ell}$ is a factor $\sim$ 20 (10) \{9\} larger than the one
required for 5$\sigma$ discovery for $M_{Z'}=500$ (1000) \{1500\} GeV
and that these factors are almost independent of $\Gamma_{Z'}/M_{Z'}$.
\smallskip

\section{Summary}

In this work we have performed a model independent analysis of the LHC
potential to unravel the spin of new charged and neutral vector
resonances associated with the EWSB sector that are predicted in many
extensions of the SM.  The production of a charged vector resonance
leads to trilepton final states and we showed that the study of the
$\cos\theta_{\ell\ell}^*$ and the reconstructed $\cos\theta_{WZ}$
distributions yield similar discriminating power for the spin of the
new state. We find that the study of the trilepton channel can lead to
a 99\% CL determination of the new charged state spin in a large fraction
of the parameter space where this state can be observed at the LHC for
an integrated luminosity of 100 fb$^{-1}$. As an illustration, let us
consider the case of Higgsless models~\cite{Csaki:2003dt} where the
$W^\prime$ mass is expected to be lighter than 1 TeV and that
\[
 \left(\frac{g_{W^\prime q\bar q}}{g_{Wq\bar q}} \frac{g_{W^\prime
        WZ}}{ {g_{W^\prime WZ}}_{max}}\right) \simeq {\cal O}(0.07)  \; .
\]
The range of these parameters indicates that this new state should be
observed with an integrated luminosity of the order of 10 fb$^{-1}$,
while the determination of its spin should require a few tens
fb$^{-1}$.

The analyses for the neutral spin--1 resonances was carried out using 
the production of opposite charge dileptons. 
It turns out that the SM background in this case can be efficiently reduced, 
however, it is not completely eliminated by the cuts, consequently leading to 
a larger
required luminosity for the discovery and determination of the spin of
the neutral new particles.  We considered two methods for the reconstruction
of the final state neutrinos (and consequently of the $Z'$ mass)  
MAOS-original and MAOS-modified, that lead to similar results. We find  that 
the variables $\cos\theta^*_{\ell\ell}$ and $\cos\theta_{WW}$ 
with $\cos\theta_{WW}$  reconstructed with any of the MAOS methods 
lead to  equivalent precision in the spin determination of the  $Z^\prime$. 
Analogously to the $W^\prime$
case, the 99\% CL determination of the neutral resonance spin can be
carried out in a large fraction of the discovery parameter space for a
fixed integrated luminosity. In the particular case of Higgsless models
~\cite{Csaki:2003dt}, the $Z^\prime$ mass is expected to be smaller
than 1 TeV and its couplings 
\[
 \left(\frac{g_{Z^\prime q\bar q}}{g_{Zq\bar q}} \frac{g_{Z^\prime
        WW}}{ {g_{Z^\prime WW}}_{max}}\right) \simeq {\cal O}(0.1) 
\]
which implies that an integrated luminosity of 100 fb$^{-1}$ should be
enough for a $Z^\prime$ discovery and to unravel its spin.

\section*{Acknowledgments}

We would like to thank E.G. Moraes and P.G. Mercadante for
enlightening discussions.  O.J.P.E is supported in part by Conselho
Nacional de Desenvolvimento Cient\'{\i}fico e Tecnol\'ogico (CNPq) and
by Funda\c{c}\~ao de Amparo \`a Pesquisa do Estado de S\~ao Paulo
(FAPESP); M.C.G-G is also supported by USA-NSF grant PHY-0969739 and
by Spanish grants from INFN-MICINN agreement program ACI2009-1038,
consolider-ingenio 2010 program CSD-2008-0037, by CUR Generalitat de
Catalunya grant 2009SGR502 and together with J.G-F by MICINN
2007-66665-C02-01. J.G-F is further supported by Spanish ME FPU grant
AP2009-2546.

\bibliographystyle{h-physrev}

\begin{thebibliography}{10}

\bibitem{Lee:1977yc}
  B.~W.~Lee, C.~Quigg, and H.~B.~Thacker,
  Phys.\ Rev.\ Lett.\  {\bf 38}, 883 (1977).

\bibitem{Lee:1977eg}
  B.~W.~Lee, C.~Quigg, and H.~B.~Thacker,
  Phys.\ Rev.\  D {\bf 16}, 1519 (1977).

\bibitem{Csaki:2003dt}
  C.~Csaki, C.~Grojean, H.~Murayama, L.~Pilo, and J.~Terning,
  Phys.\ Rev.\  D {\bf 69}, 055006 (2004)
  [arXiv:hep-ph/0305237];
  C.~Csaki, C.~Grojean, L.~Pilo, and J.~Terning,
  Phys.\ Rev.\ Lett.\  {\bf 92}, 101802 (2004)
  [arXiv:hep-ph/0308038];
  G.~Cacciapaglia, C.~Csaki, C.~Grojean, and J.~Terning, 
  Phys.\ Rev.\ D {\bf 70}, 075014 (2004) 
  [arXiv:hep-ph/0401160];
  G.~Cacciapaglia, C.~Csaki, G.~Marandella, and J.~Terning,
  Phys.\ Rev.\  D {\bf 75}, 015003 (2007)
  [arXiv:hep-ph/0607146];
  Y.~Nomura,
  J.\ High Energy Phys. {\bf 0311}, 050 (2003)
  [arXiv:hep-ph/0309189];
  R.~S.~Chivukula, D.~A.~Dicus, H.~-J.~He,
  Phys.\ Lett.\  {\bf B525}, 175-182 (2002).
  [hep-ph/0111016];
  R.~S.~Chivukula, H.~-J.~He,
  Phys.\ Lett.\  {\bf B532}, 121-128 (2002).
  [hep-ph/0201164].


\bibitem{TC}
S.~Dimopoulos and L.~Susskind,
Nucl.\ Phys.\ B {\bf 155}, 237 (1979);
L.~Susskind,
Phys.\ Rev.\ D {\bf 20}, 2619 (1979);
S.~Weinberg,
Phys.\ Rev.\ D {\bf 19}, 1277 (1979).

\bibitem{NTC} See for instance,
  C.~T.~Hill and E.~H.~Simmons,
  Phys.\ Rept.\  {\bf 381}, 235 (2003)
  [Erratum-ibid.\  {\bf 390}, 553 (2004)]
  [arXiv:hep-ph/0203079];


\bibitem{Birkedal:2004au}
A.~Birkedal, K.~Matchev, and M.~Perelstein,
Phys.\ Rev.\ Lett.\  {\bf 94}, 191803 (2005)
[arXiv:hep-ph/0412278].


\bibitem{Alves:2008up}
  A.~Alves, O.~J.~P.~Eboli, M.~C.~Gonzalez-Garcia {\it et al.},
  Phys.\ Rev.\  {\bf D79}, 035009 (2009).
  [arXiv:0810.1952 [hep-ph]].


\bibitem{Asano:2010ii} The $Z^\prime$  leptonic decay can be observed
in its associated production with a hadronically decaying $W$, see
  M.~Asano, Y.~Shimizu,
  JHEP {\bf 1101}, 124 (2011).
  [arXiv:1010.5230 [hep-ph]].


\bibitem{susyspin1}
A.~J.~Barr,
Phys.\ Lett.\ B {\bf 596}, 205 (2004)
[arXiv:hep-ph/0405052].
 A.~J.~Barr,
J.\ High Energy Phys. {\bf 0602}, 042 (2006) 
[arXiv:hep-ph/0511115].

\bibitem{susyspin2}
J.~M.~Smillie and B.~R.~Webber,
J.\ High Energy Phys. {\bf 0510}, 069 (2005),
[arXiv:hep-ph/0507170];
A.~Alves, O.~\'Eboli, and T.~Plehn,
Phys.\ Rev.\  D {\bf 74}, 095010 (2006)
[arXiv:hep-ph/0605067];
A.~Alves and O.~\'Eboli,
Phys.\ Rev.\  D {\bf 75}, 115013 (2007)
[arXiv:0704.0254 [hep-ph]].



\bibitem{spinreferences} 
 R.~Cousins, J.~Mumford, J.~Tucker {\it et al.},
  JHEP {\bf 0511}, 046 (2005);
 Y.~Gao,  Y.~Gao, A.~V.~Gritsan, Z.~Guo {\it et al.},
  Phys.\ Rev.\  {\bf D81}, 075022 (2010).
  [arXiv:1001.3396 [hep-ph]];
 C.~Englert, C.~Hackstein, M.~Spannowsky,
  Phys.\ Rev.\  {\bf D82}, 114024 (2010).
  [arXiv:1010.0676 [hep-ph]];
M.~R.~Buckley, H.~Murayama, W.~Klemm, and V.~Rentala,
Phys.\ Rev.\  D {\bf 78}, 014028 (2008)
[arXiv:0711.0364 [hep-ph]];
  L.~T.~Wang and I.~Yavin,
  JHEP {\bf 0704}, 032 (2007)
  [arXiv:hep-ph/0605296];
  L.~Edelhauser, W.~Porod, R.~K.~Singh,
  JHEP {\bf 1008}, 053 (2010).
  [arXiv:1005.3720 [hep-ph]];
 W.~Ehrenfeld, A.~Freitas, A.~Landwehr {\it et al.},
  JHEP {\bf 0907}, 056 (2009).
  [arXiv:0904.1293 [hep-ph]];
F.~Boudjema, R.~K.~Singh,
  JHEP {\bf 0907}, 028 (2009).
  [arXiv:0903.4705 [hep-ph]];
O.~Gedalia, S.~J.~Lee, G.~Perez,
  Phys.\ Rev.\  {\bf D80}, 035012 (2009).
  [arXiv:0901.4438 [hep-ph]];
L.~-T.~Wang, I.~Yavin,
  Int.\ J.\ Mod.\ Phys.\  {\bf A23}, 4647-4668 (2008).
  [arXiv:0802.2726 [hep-ph]].




\bibitem{Alves:2009aa}
  A.~Alves, O.~J.~P.~Eboli, D.~Goncalves {\it et al.},
  Phys.\ Rev.\  {\bf D80}, 073011 (2009).
  [arXiv:0907.2915 [hep-ph]].

\bibitem{maos}
  W.~S.~Cho, K.~Choi, Y.~G.~Kim {\it et al.},
  Phys.\ Rev.\  {\bf D79}, 031701 (2009).
  [arXiv:0810.4853 [hep-ph]].

\bibitem{madevent}
T.~Stelzer and F.~Long,
Comput.{} Phys.{} Commun.{} \textbf{81} (1994) 357;
F.~Maltoni and T.~Stelzer,
J.\ High Energy Phys. {\bf 0302}, 027 (2003)
[arXiv:hep-ph/0208156].

\bibitem{CTEQ6}
J.~Pumplin, D.~R.~Stump, J.~Huston, H.~L.~Lai, P.~Nadolsky and W.~K.~Tung,
  JHEP {\bf 0207}, 012 (2002)
  [arXiv:hep-ph/0201195].

\bibitem{cdf}
T.~Aaltonen {\it et al.} [ The CDF Collaboration ],
  Phys.\ Rev.\ Lett.\  {\bf 104}, 241801 (2010).
  [arXiv:1004.4946 [hep-ex]].

\bibitem{Lester:1999tx}
  C.~G.~Lester, D.~J.~Summers,
  Phys.\ Lett.\  {\bf B463}, 99-103 (1999).
  [hep-ph/9906349].

\bibitem{maos2}
K.~Choi, S.~Choi, J.~S.~Lee {\it et al.},
  Phys.\ Rev.\  {\bf D80}, 073010 (2009).
  [arXiv:0908.0079 [hep-ph]].

\bibitem{rainth}
D.~Rainwater, Ph.D. thesis, report arXiv:hep-ph/9908378.
%

\bibitem{atlas}
V.~Cavasinni, D.~Costanzo and I.~Vivarelli, ATL-PHYS-2002-008;
S.~Asai {\it et al.},
Eur.\ Phys.\ J.\ C {\bf 32S2}, 19 (2004).




\end{thebibliography}

\end{document}